%% file: sample-sigconf.tex
\documentclass[sigconf]{acmart}
\usepackage{booktabs} % For formal tables

\setcopyright{none}
\acmDOI{}

\acmISBN{}

\acmConference[]{}{}{}

\usepackage[ruled,vlined]{algorithm2e}

\begin{document}
\title{Event-Radar: Real-time Local Event Detection System for Geo-Tagged Tweet Streams}
%\titlenote{Produces the permission block, and
%  copyright information}
%\subtitle{Extended Abstract}
%\subtitlenote{The full version of the author's guide is available as
%  \texttt{acmart.pdf} document}

\author{Sibo Zhang}
%\authornote{Dr.~Trovato insisted his name be first.}
\affiliation{%
  \institution{University of Illinois at Urbana-Champaign}
%  \streetaddress{P.O. Box 1212}
}
\email{siboz2@illinois.edu}

\author{Yuan Cheng}
%\authornote{The secretary disavows any knowledge of this author's actions.}
\affiliation{%
  \institution{University of Illinois at Urbana-Champaign}
%  \streetaddress{P.O. Box 1212}
}
\email{yuanc3@illinois.edu}

\author{Deyuan Ke}
%\authornote{This author is the
%  one who did all the really hard work.}
\affiliation{%
  \institution{University of Illinois at Urbana-Champaign}
%  \streetaddress{1 Th{\o}rv{\"a}ld Circle}
%  \country{Iceland}
}
\email{deyuank2@illinois.edu}

\renewcommand{\shortauthors}{Sibo Zhang et al.}

\begin{abstract}
The local event detection is to use people's posting messages with geotags
on social networks to reveal the related ongoing events and their locations \cite{1}. Recent studies have demonstrated that the geo-tagged tweet stream
serves as an unprecedentedly valuable source for local event detection.
Nevertheless, how to effectively extract local events from large geo-tagged
tweet streams in real time remains challenging. A robust and efficient
cloud-based real-time local event detection software system would benefit
various aspects in the real-life society, from shopping recommendation for
customer service providers to disaster alarming for emergency departments.

We use the preliminary research GeoBurst \cite{1} as a starting point,
which proposed a novel method to detect local events. {GeoBurst+} \cite{2}
leverages a novel cross-modal authority measure to identify several pivots
in the query window. Such pivots reveal different geo-topical activities and
naturally attract related tweets to form candidate events. It further
summarises the continuous stream and compares the candidates against the
historical summaries to pinpoint truly interesting local events. We mainly
implement a website demonstration system Event-Radar with an improved
algorithm to show the real-time local events online for public interests.
Better still, as the query window shifts, our method can update the event
list with little time cost, thus achieving continuous monitoring of the
stream.
\end{abstract}


%\ccsdesc[500]{Computer systems organization~Embedded systems}


\maketitle

\input{samplebody-conf}

\bibliographystyle{ACM-Reference-Format}
%\bibliography{sample-bibliography}

\end{document}

%% file: samplebody-conf.tex
\section{Introduction}

With over 500 million tweets written by users every day and there are more 
than 100 million users, Twitter has been one of the most popular online news 
and social networking service. This means that a large of amount of data is 
frequently generated. Users post and interact with tweets, which restricted 
to 140 characters. Beyond Twitter, we have online social media sites like 
Facebook, Youtube and Instagram, which have transformed the method we 
connect with individuals, groups, and communities and altered everyday 
practices \cite{5}. Numerous recent workshops, such as Semantic Analysis in 
Social Media \cite{6}, are increasingly focusing on the influence of social media 
on our daily lives. Unlike other media sources, Twitter messages offer 
timely and fine-grained information about any event, reflecting personal 
perspectives, social information, emotional reactions, and local event.

A local event is an unusual activity burst in a local area and 
within specified duration while engaging a considerable number of 
participants. Empirical studies \cite{7} show that the online social networking 
service Twitter is often the first medium to break significant natural events 
such as earthquakes often in a matter of seconds after they occur. Twitter 
is ``what's-happening-right-now'' tool \cite{8} and given the nature of its 
tweets are a real-time flow of text messages coming from very different 
sources covering various kinds of subjects in distinct languages and 
locations. The Twitter free stream is an interesting source of data for 
``real time'' event detection based on text mining techniques. Noticing that 
here ``real time'' means that events need to be discovered as early as 
possible after they start unravelling in the online social networking 
service stream. Such information about emerging events can be hugely 
valuable if it is visible in real time.

Studying those data can provide us with useful information. ``What 
is happening right now?'' is a fascinating question that many people ask 
every day. People are interested in those events happens locally \cite{9}. 
Corporations are interested in sponsoring their product to favourable 
customers \cite{10}. Event detection can answer this question. Besides that, 
nature disasters might be detected by Twitter and warn people even faster 
than other media \cite{11}. Some predictions can also be completed from Twitter 
data, such as the crime prediction \cite{12}. Typical examples include the bomb 
blasts in Mumbai in November 2008, the flooding of the Red River Valley in 
the United States and Canada in March and April 2009, and the ``Arab 
Spring'' in the Middle East and North Africa region \cite{13}. Several studies 
have analysed Twitter's user intentions. For example, user intentions can be 
categorised on Twitter into daily chatter, conversations, allocation 
information, and journalism news. They also identified Twitter users as 
information sources, friends, and information hunters.

Considerable research efforts have been made in detecting real-time events. 
However, most them lack the accuracy when dealing with local events or the 
capability for real-time events. For example, Abdelhaq's EvenTweet \cite{3}, 
using spatial entropy, clustering, and feature ranking to extract and rank 
local events, cannot deal with the real-time environment.

Nevertheless, there are also trials of event detection in Twitter 
rivalling to event detection in traditional media. Twitter messages are 
usually not well organised. Twitter streams cover huge amounts of 
meaningless messages, which negatively affect the detection performance. 
Furthermore, conventional text mining techniques are not appropriate, since 
the short length of tweets, a significant number of spelling and grammatical 
mistakes, and the chronic use of straightforward and mixed language. Since 
spelling and grammar errors, mixed languages, colloquial expressions and 
shortened words are very common in tweets, we are very hard to understand 
their semantic meanings. Similarly, while a significant amount of data, it 
is tough to find a well-organized way to select valuable tweets in Twitter. 
While the real-time detection of local events was nearly incredible years 
ago due to the lack of reliable data sources, the explosive growth of 
geo-tagged tweet data brings new opportunities to it. With the ubiquitous 
connectivity of wireless networks and the vast proliferation of mobile 
devices, more than 10 million geo-tagged tweets are created in the Twitter 
every day. Numerous real-world examples have exposed the effectiveness and 
the timely information reported by Twitter during disasters and social 
movements. For example, when the Tohoku Earthquake hit Japan on March 2011 
and when the Baltimore Riot took place in April 2015, many people posted 
geotagged tweets to broadcast it right there. Its sheer size, multi-faceted 
information, and real-time nature make the geo-tagged tweet stream an 
unprecedentedly valuable source for detecting local events \cite{2}.

Tweets are about contents from daily life things to newest local and 
worldwide events. Twitter streams contain significant amounts of meaningless 
messages (pointless babbles) and rumours \cite{13}. These are important to help 
to understand people's reactions to events. Nevertheless, they undesirably 
affect event detection performance.

A major test facing event detection from Twitter streams is to 
separate the dull and polluted information from exciting real-life events. 
In practice, highly scalable and efficient approaches are required for 
handling and processing the increasingly significant quantity of Twitter 
data especially for real-time event detection. Other challenges are 
intrinsic to Twitter's natural. These are due to the short length of tweet 
messages, the frequent use of simple words, the enormous quantity of 
spelling and grammatical errors. Such data sparseness, lack of context, and 
diversity of vocabulary make the traditional text analysis techniques less 
appropriate for tweets \cite{14}. Also, different events may enjoy different 
popularity among users and can differ significantly in content, the number 
of messages and participants, periods, internal structure, and causal 
relationships \cite{15}.

Thus, the challenges are in below three aspects:

\textbf{1. Integrating diverse types of data.} The geo-tagged tweet 
stream involves three different data types: location, time, and text. 
Considering the entirely different representations of those data types and 
the complex cross-modal interactions among them, how to effectively 
integrate them for local event detection is challenging.

\textbf{2. Capturing the semantics of short text.} Since every tweet 
is limited to 140 characters, the semantics of the user's activity is 
expressed through short and sparse text messages. Compared with traditional 
documents (\textit{e.g.}, news), it is much harder to capture the semantics of short 
tweet messages and extract high-quality local events.

\textbf{3. On-line and real-time detection.} When a local event 
outbreaks, it is key to report the event instantly to allow for timely 
actions. As massive geo-tagged tweets stream in, the detector should work in 
an on-line and real-time manner instead of a batch-wise and inefficient one. 
Such a requirement is the third challenge of our problem \cite{2}.

\section{RELATED WORK}
The Topic Detection and Tracking program by Jonathan G. Fiscus and George R. 
Doddington \cite{16} gave the following definitions of the event:

\textbf{Event} is ``something that happens at some specific time and place 
along with all necessary preconditions and unavoidable consequences'';

Sakaki et al. \cite{17} defines an event as an arbitrary classification of 
space/time region that might have actively participating agents, passive 
factors, products, and a location in space/time like is being defined in the 
event ontology by Raimond and Abdallah \cite{18}. The target events in this work 
are significant events that are visible through messages, posts, or status 
updates of active users in Twitter online social network service. These 
events have several properties: (i) they are of large scale because many 
users experience the event, (ii) they particularly influence people's daily 
life, being that the main reason why users are induced to mention it, and 
(iii) they have both spatial and temporal regions. The importance of an 
event is connected with the distance users have between themselves and the 
event, and with the spent time since the occurrence.
\begin{figure}[!htb]
	\includegraphics[width=0.75\linewidth]{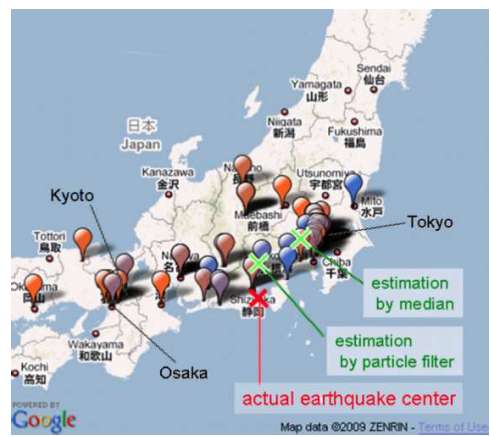}
	\caption{Earthquake location estimation based on tweets. Balloons show the tweets on the earthquake. The cross shows the earthquake centre. Red represents new tweets; blue represents later tweets \cite{17}.}
\end{figure}
\begin{figure}
	\includegraphics[width=0.75\linewidth]{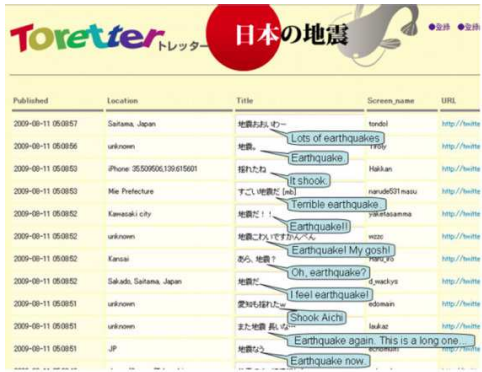}
	\caption{Screenshot of Trotter, an earthquake reporting system \cite{17}.}
\end{figure}

\subsection{Detection Task}

Events are evaluated using a decision based on whether a document reports a 
new topic that has not been reported previously, or if should be merged with 
an existent event \cite{22}. Differing on how data is treated, two groups of 
Event Detection systems were identified \cite{23}.

\textbf{Online New Event Detection (NED).} 

Online New Event Detection denotes to the task of classifying events from 
live streams of tweets in real-time. Most new and retrospective event 
detection methods rely on the use of well-known clustering-based algorithms 
\cite{24}. Usually, new event detection contains the continuous monitoring of 
tweet feeds for discovering events in near real time, which could do event 
detection of real-world events like breaking news, natural disasters or 
football game.

\textbf{Retrospective Event Detection (RED).} 

Retrospective Event Detection denotes to the process of classifying 
unidentified events previously from gathered past data that have arrived in 
the past. In Retrospective Event Detection, most methods are founded on the 
retrieval of event relevant documents by performing queries over a 
collection of records. Both techniques assume that event relevant documents 
contain the query terms. A disparity of the previous approach is the use of 
query growth techniques, meaning that some messages related to a specific 
event do not contain specific event related information, but with the use of 
improved queries, messages related to the event can be recovered.

\subsection{Type of Event}

Event detection could be classified into specified or unspecified event 
detection techniques \cite{25}. By using specific pre-known information and 
features about an event, traditional information retrieval and extraction 
techniques can be modified to perform specified event detection. Most 
traditional information retrieval and extraction methods are useless when no 
previous information is available about the event. Unspecified event 
detection methods address the issue on the basis that temporal signals 
constructed via document analysis can detect real work events. Monitoring 
trends in text streams, alliance topographies with same viewpoints, and 
categorising events into different categories are among those tasks to 
perform unspecified event detection.

\subsection{Event Detection Overview}

Event detection has been deeply studied in the past few years, and various 
methods have been proposed to address the problem. Frequently used feature 
representations are also presented and discussed. This survey does not 
provide an exhaustive review of existing approaches but rather techniques 
which related to the area that would focus on our most important research 
directions.

The event detection problem is not a new research topic, Yang et al. 
\cite{26} in 1998, is a study on retrospective and on-line event detection which 
examined the usage and postponement of text retrieval and clustering 
techniques. The main task was to detect new events from a well-organized 
stream of news stories repeatedly. The system performed quite well and 
showed that basic techniques such as document clustering could be highly 
effective to perform event detection. Depending on the type of events, these 
methods are classified into unspecified and specified event detection.

\textbf{Unspecified Event Detection: }This kind of events is mainly about 
emerging events, breaking news, and general topics that attract a 
considerable number of users' attentions. We are interested in using Twitter 
tweets to find ongoing local events. Thus, the general events will be of our 
interests \cite{34}. Typically, such events often come with a 
significant temporary boost of the use of keywords. The trends in tweets can 
be clustered according to the frequent-occur feature. Whereas, there is a 
non-toxic event which is viewed as noise when conducting event detection 
\cite{27}. Therefore, the major challenge to be dealt with in unspecified event 
detections is to distinguish significant trends event from those trivial 
non-events. Several techniques have been proposed to tackle this challenge 
by applying a range of machine learning, data mining, and text mining 
techniques.

\textbf{TwitterStand:} News in tweets \cite{35} showed a novel system which deals 
with the problem of capturing proper tweets trends related to breaking news, 
TwitterStand. Two techniques were used, a na\"{\i}ve Bayes classifier and 
online clustering algorithms. Na\"{\i}ve Bayes classifier was applied to 
distinguish breaking news from irrelevant non-events in tweet streaming. 
Whereas, the online cluster which employs term frequency--inverse document 
frequency and cos similarity measures were used to from newsgroups. The 
paper used tweets' hashtag and timestamps as an additional method to reduce 
the clustering errors of online cluster algorithms.

\textbf{Breaking news detection} and tracking on Twitter \cite{36} proposed a 
technique to capture breaking news from Twitter, with the additional 
functions of following and ranking. The tweets were extracted through 
pre-defined queries of Twitter API and indexed before similarity grouping. 
Grouping was based on the of-if similarity measure between messages. All 
tweets were sorted in ascending weight with authors, proper nouns, and 
hashtags. Stanford Named Entity Recognizer (NER) was used to identify proper 
nouns with the number of sponsors' followers and some shares of the tweets 
taken into consideration. In the paper, the author factor was introduced for 
the reliance and soundness of the tweets, which improved the accuracy. They 
also developed an application called Hot-streams to validate the algorithms.

Streaming first story detection with application to Twitter \cite{37} focused on predicting new events which never occurred in previous tweets. 
The approach was mainly about improving the efficiency of conducting cosine 
similarity measurement within documents. The paper developed the locality 
sensitive hashing methods, which applied the search operations to a small 
number of records and optimised the complexity within a constant time and 
space. Whereas, the replies, the number of shares, and hashtags were not 
taken into consideration in the paper. The experiment results indicated two 
remarkable facts which are 1. User based prevails when compared with 
tweet-based ranking. 2. The entropy of information leads to less message 
spam.

\textbf{Real-world event identification on Twitter} \cite{38} used an online 
clustering technique to associate tweets with the real-world event. It keeps 
clustering related tweets and then classifies the clusters into two 
categories, real events or trivial nonevents. A significant difference 
between actual events and nonevents is that there are Twitter-centric topics 
within nonevents. Naaman pointed out that such topic is trending but do not 
reflect or represent any real-world events. All tweets were present as an 
of-if weight vector based on their contents. The paper used cosine 
similarity to calculate the distance between a tweet and cluster centroids. 
The weight of hashtag was doubled as it was hypothesized that it brings a 
strong connection between text and tweet topic. All standard methods of 
pre-processing, such as stem and stop word list were also used. The cluster 
was computed with a combination of term-frequency-based temporal features, 
twitter-operation-based social functions, local features, and 
Twitter-centric features. Term {frequency-based} features based on the number 
of the appearance in the message set with a cluster. The twitter operations 
include comments, replies, share, etc, and the feature contains the 
percentage of those operations in the cluster content. This paper assumes 
that the proposed cluster obtained the intention to revolve around a certain 
meaningful topic but the non-event clusters have the trends of revolving 
around some irrelevant terms such as ``dinner'', ``sleep'', or ``right''. 
The twitter-centric. Upon all these work, an SVM is developed to classify 
the clusters and tweets associated with the clusters into real-world 
labelled portion or non-even labelled portion.

Towards effective event detection, tracking and summarization on 
microblog data \cite{39} proposed a technique to assign some topic-word-based 
features to the microblog data to train a cluster. Topic words are those 
words which share more popularity when compared with others in an event. 
These words are computed from an extraction of daily messages in microblog 
data based on the frequency of the phrase, incidence of hashtag associated 
with the phrase, and entropy. A co-occurrence graph was generated by adding 
edges to messages and topical words where a hierarchical cluster was used 
upon to transfer the set of topical words into event clusters. The paper 
claims that the hierarchical cluster over forms traditional K-means 
algorithms.

Real-time event detection for online behavioural analysis of big 
social data, \cite{40} employed a 5-stage method in real-time event detection. It 
collected tweets with search conditions and converted them into JSON format. 
The terms are then extracted from the tweets by adopting named entity 
recognition. It constructs the signals by tracking both occurrences of terms 
extracted from extraction phase and diffusion of the information. After 
this, a weighted graph of which nodes are tweets is computed. The edges are 
measured by the complement of the similarity degree. Clustering is a final 
stage which includes adjacent points that are close measured by timestamp 
and occurrences. Each cluster is viewed as a potential candidate for 
grouping events to whether they are real-world events or non-events.

\textbf{Specified Event Detection}: A specified event can be public or 
pre-planned social meetings such as a concert. It should contain the 
metadata such as venue, time, attendees, and musicians. The work introduced 
here attempt to exploit Twitter textual content or metadata information or 
both.

Popescu and Pennacchiotti \cite{41} focused on identifying controversial 
events that provoke public discussions with opposing opinions on Twitter, 
such as controversies involving superstars. Their detection outline is based 
on the idea of a Twitter snapshot, a trio consisting of a target entity, a 
given period, and a set of tweets about the entity from the target period. 
Assumed a set of Twitter snapshots, an event detection module first 
distinguishes between the event and non-event snaps using a supervised 
gradient boosted decision trees \cite{42}, trained on the manually labelled data 
set. To rank these event snaps, a controversy model allocates higher scores 
to controversial event snapshots, by a reversion algorithm applied to a 
large number of features. The employed features are based on 
Twitter-specific characteristics including linguistic, structural, 
buzziness, nine sentiment, and controversy features, and on external 
features for example news buzz. These external features require time 
alignment of entities in news media and Twitter sources, to capture entities 
that are trending in both sources because they are more likely to mention 
real-world events. The authors have also planned to merge the two stages 
into a single-stage system by including the event detection score as an 
extra feature into the controversy model. This produced an improved 
performance. Feature analysis of the single-stage system exposed that the 
event score is the most relevant feature because it discriminates event from 
nonevent snapshots. Hashtags are originated to be important semantic 
topographies for tweets in the meantime they help classify the topic of a 
tweet and approximation the topical cohesiveness of a set of tweets. 
External features based on news and the Web are also originated usefully; 
hereafter, association with traditional media helps authenticate and explain 
social media reactions. Also, the linguistic, structural, and sentiment 
features also deliver significant effects. The authors determined that a 
rich, diverse set of features be crucial for controversy detection.

Benson et al. \cite{43} present a novel approach to identify Twitter 
messages for concert events using a factor graph model, which simultaneously 
examines individual messages, clusters them according to the event type, and 
induces a correct value for each event property. The motivation is to infer 
a comprehensive list of musical events from Twitter (based on artist--venue 
pairs) to whole an existing list (\textit{e.g.}, city event calendar table) by 
discovering new musical events mentioned by Twitter users that are difficult 
to find in other media sources. At the message level, this approach relies 
on a conditional random field (CRF) to excerpt the artist name and position 
of the event. The contribution features to CRF model include word form; a 
set of even expressions for mutual emoticons, time references, and venue 
types; a large number of words for artist names removed from an external 
source; and a bag of words for city place names. Clustering is directed by 
term popularity, which is an arrangement score among the message term labels 
and some candidate worth. To imprisonment the huge text difference in 
Twitter messages, this score is founded on a weighted combination of term 
similarity measures. This including complete string matching, and adjacency 
and equality indicators scaled by the inverse document frequency. Also, a 
uniqueness factor is working during clustering to expose rare event messages 
that are dominated by the general ones and to discourage various messages 
from the same facts to cluster into multiple incidents. Alternatively, a 
consistent indicator is employed to discourage messages from multiple events 
to form a single cluster. The factor graph model is then used to capture the 
interaction between all components and provide the final choice. The 
production of the model consists of an event-based clustering of messages, 
where each cluster is characterised by an artist--venue pairs.

Lee and Sumiya \cite{44} present a geosocial local event detection system 
based on modelling and monitoring crowd behaviours via Twitter, to identify 
local festivals. They rely on geographical regularities deduced from the 
usual behaviour patterns of crowds using geotags. First, Twitter geotagged 
data are collected and preprocessed over an extended period for a specific 
region \cite{45}. The area is then alienated into several regions of interest 
(ROI) using the k-means algorithm, applied to the geographical coordinates 
(longitudes/latitudes) of the collected data. Geographical regularities of 
the crowd within each ROI are then predictable from historical data based on 
three main features: some tweets, users, and moving users within an ROI. 
Statistics for these functions are then accumulated over historical data 
using 6-hour time interval to form the estimated behaviour of the crowd 
within each ROI. Finally, unusual events in the monitored geographical area 
can be detected by comparing statistics from new tweets with those of the 
estimated behaviour. The authors found that an augmented user combined with 
an increased number of tweets provides a strong indicator of local 
festivals.

\subsection{Local Event Detection}

Abdelhaq \cite{31} presents a method of EVENTTWEET which extracts hashtags and 
Twitter keywords based on temporal burst and spatial location. Then it 
employs a cluster on these keywords to compute events depending on location 
distribution.

Krumm, John \cite{30} introduced a method, Eyewitness, to find local 
events from a large-scale stream of Twitter textures. The paper considers 
the location statistics of tweets and classifies the location data to train 
a classifier for locating meaningful tweets. Then it envisions the 
classifier being harnessed in a user-interaction system to identify and 
monitor the events based on users' locations. Eyewitness adopts a regression 
model to predict the number of geotagged tweets in a certain amount of time. 
If the real number of tweets is larger than the predicted number, the event 
is defined as a local event. It also employs a text summarization algorithm 
to extract the tweets belongs to the event. 

Chao \cite{1} proposed another approach, GEOBURST for local event 
detection. The paper assumed that a significant local event results in the 
scene of many geotagged texts around one certain place. Moreover, a method 
was built based on this assumption which firstly looks for all geo-clustered 
topics and secondly ranks those topics based on spatiotemporal business to 
get the significant local events. There is an ad-hoc streaming process 
embedded in the methods to implement the function of processing and updating 
continuous real-time tweets.

\section{PRELIMINARIES}

In this section, we describe the application of Local Event Detection 
algorithm, which consisted of Candidate Generator, Candidate Generator 
Classification, Online Updater. 

\subsection{Candidate Generator}

Given a query window Q and the set $ D_Q $ of tweets falling in Q, the candidate 
generator is to divide $ D_Q $ into several geo-topical clusters, such that the 
tweets in each group are geographically close and semantically coherent. The
Clustering of $ D_Q $, however, poses several challenges: how to combine the 
geographical and semantic similarities in a reasonable way? How to capture 
the correlations between different keywords? Moreover, how to generate 
quality clusters without knowing the suitable number of clusters in advance? 
To address these challenges, we perform a novel pivot seeking process to 
identify the centres of geo-topical clusters. Our key insight is that: the 
spot where the event occurs acts as a pivot that produces relevant tweets 
around it; the closer we are to the pivot, the more likely we observe 
relevant tweets. Therefore, we define a geo-topical authority score for each 
tweet, where a kernel function captures the geographical influence among 
tweets, and the semantic influenced by random walk on a keyword 
co-occurrence graph. With this authority measure, we develop an authority 
ascent procedure to retrieve authority maxima as pivots; and each pivot 
naturally attracts similar tweets to form a quality geo-topical cluster. 
Below, we rest introduce our geo-topical authority measure to define pivot 
tweets and then develop an authority ascent procedure for pivot seeking.

\subsubsection{Pivot Tweet}

Pivot Tweet is an amount of $ G (d_0 \to d_1) $ energy is distributed from $ d_0 $ to $ d $ 
through random walk on the graph, $ G(d_0 \to d_1) $  $ S (d_0 !d) $ is the amount 
that successfully reaches $ d $; and $ d_0 $ authority is the total sum of energy 
that d receives from its neighbors \cite{38}. The authority score is analogous to 
kernel density in the task of nonparametric kernel density estimation \cite{7}. 
In kernel density estimation, the density of any point x in the Euclidean 
space is contributed mainly by the observed points that are close enough to 
x. As such, the density maxima can be defined in a
non-parametric manner. Analogously, in our problem, the geo-topic authority 
of any tweet d is contributed by the observed tweets that are similar to $ d $ 
both geographically and semantically. As a result, the salient tweets for 
different activities can be selected in the geo-topical space.
\begin{algorithm}
	\caption{Pivot seeking.}
	\KwIn{The tweet set $ D_Q $, the kernel bandwidth $ h $, the semantic threshold $ \delta $.}
	\KwOut{The pivot for each tweet in $ D_Q $.}	\tcp{Neighborhood computation.}
	\ForEach{$ d \in D_Q $}{$ N(d) \leftarrow \{d' | d' \in D_Q, G(d' \to d) > 0, S(d' \to d) > \delta \} $\;}
	\tcp{Authority computation.}
	\ForEach{$ d \in D_Q $}{$ A(d) \leftarrow d $'s authorith score computed from $ N(d) $\;}
	\tcp{Find local pivot for each tweet.}
	\For{$ d \in D_Q $}{$ l(d) \leftarrow \arg \max\limits_{d' \in N(d)} A(d') $\;}
	\tcp{Authority ascent.}
	\ForEach{$ d \in D_Q $}{Perform authority ascent to find the pivot for $ d $\;}
\end{algorithm}

\begin{algorithm}
	\caption{Approximate RWR score computation.}
	\KwIn{The keyword co-occurrence graph $ G $, a keyword $ q $, the restart probability $ \alpha $, an error bound $ \epsilon $.}
	\KwOut{$ q $'s vicinity $ V_q $.}
	\tcp{$ p(u) $ is the score of node $ u $ that needs to be propagated.}
	$ s(q) \leftarrow \alpha, p(q) \leftarrow \alpha, V_q \leftarrow \phi $\;
	$ Q \leftarrow $ a priority queue that keeps $ p(u) $ for the keywords in $ G $\;
	\While{$ Q.peek() \geq \alpha\phi $}{
	$ u \leftarrow  $Q.pop()\;
	\For{$ v \in I(u) $}{
		$ \Delta s(v) = (1 - \alpha) p_{vu}p(u) $\;
		$ s(v) \leftarrow s(v) + \Delta s(v) $\;
		$ V_q[v] \leftarrow s(v) $\;
		Q.update$ (v, p(v) + \Delta s(v)) $\;}
	$ p(u) \leftarrow 0 $\;
	}
	\Return $ V_q $\;
\end{algorithm}

\subsubsection{Authority Ascent for Detecting Geo-Topical Clusters}

Now our task is to nd all pivots in $ D_Q $ and assign each tweet to its 
corresponding pivot. We develop an authority ascent procedure for this 
purpose. As shown in Figure 3, starting from a tweet $ d_1 $ as the initial 
center, we perform step-by-step center shifting. Assuming the center at step 
$ t $ is tweet $ dt $, we nd $ dt $ neighborhood $ N (dt) $, and the local pivot $ l (dt) $? the tweet having the largest authority in $ N (dt) $. Then we regard $ l (dt) $ 
as our new center, i.e., $dt+1 = l (dt)$. As we continue such an 
authority ascent process, the center is guaranteed to converge to an 
authority maximum. It is because every shift operation increases the 
authority of the curr

\begin{figure}
	\includegraphics[width=\linewidth]{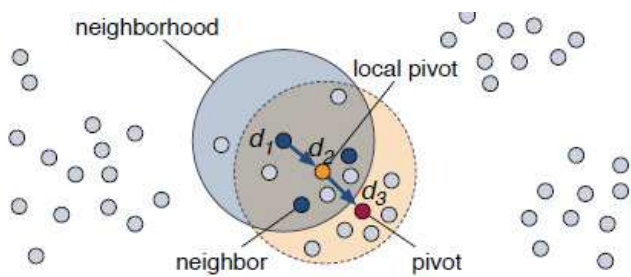}
	\caption{An illustration of the authority ascent process.}
\end{figure}

\subsection{Candidate Generator Classification}

Up to now, we have obtained a set of geo-topical clusters in the query 
window as candidate events. Nevertheless, as aforementioned, not necessarily 
does every candidate correspond to a local event. In this section, we 
describe the module for candidate event classification. The foundation of 
our classification is the summarization module, which learns word embedding 
to capture the semantics of short tweet messages and meanwhile constructs 
the activity timeline to reveal routine regional activities. In what 
follows, we describe embedding learning and activity timeline construction 
and then present the classier.

\subsubsection{Learning Embeddings from the Stream}

The embedding learner aims at capturing the semantics of short text by 
jointly mapping the tweet messages and keywords into the same 
low-dimensional space. If two tweets (keywords) are semantically similar, 
they are forced to have close embedding vectors in the latent space. The 
learner continuously consumes a massive amount of tweets from the input 
stream and learns to preserve their intrinsic semantics. As such, it can 
generate red-length vectors for any text pieces (\textit{e.g.}, the candidate event 
and the background activity), which serve as high-quality features to
discriminate whether a candidate event is indeed a local event or 
not.Relying on the tweet caching strategy and the SGD optimisation 
procedure, the embedding learner continuously consumes the geo-tagged tweet 
stream and keeps updating the embeddings for different keywords and tweets. 
With the learnt keyword embeddings, the embedding of any ad-hoc text piece 
can be easily derived with SGD. As we will illustrate shortly, such a 
property enables us to quantify the spatiotemporal unusualness of each 
candidate event and extract highly discriminative features to pinpoint true 
local events.

\subsubsection{Activity Timeline Construction}

The activity timeline aims at unveiling the normal activities in different 
regions during different time periods. For this purpose, we design a 
structure called tweet cluster (TC) and extend the CluStream algorithm 
\cite{2}.The TC essentially provides a concise where-when-what summary for S: (1) 
where: with n, ml, and 2, one can easily compute the location mean and 
variance for S; (2) when: with n, mt, and Mt 2, one can compute the average 
time and temporal variance for S; and (3) what: me keeps the number of 
occurrences for each keyword. These fields in a TC-S enable us to estimate 
the number of keyword occurrences at any location. First, the quantities n, 
ml, and ml 2 allow us to compute the center location of the TC S. Second, 
theme tracks the number of occurrences for different keywords around the 
centered location of S. With either spatial interpolation or kernel density 
estimation, one can estimate the occurrences of keyword k at any ad-hoc 
location based on the distance to the center location of S. Moreover, TC 
satisfies the additive property, i.e., the fields can be easily incremented 
if a new tweet is absorbed. Based on this property, we adapt CluStream to 
continuously clusters the stream into a set of TCs. When a new tweet $ d $ 
arrives, it ends the TCM that is geographically closest to d. If $ d $ is within 
$ M $'s boundary (computed from n, ml, and 2, see \cite{2} for details), it absorbs 
d into and updates its fields; otherwise, it creates a new TC for d. 
Meanwhile, we employ two strategies to limit the maximum number of TCs: (1) 
deleting the TCs that are too old and contain few tweets; 
Moreover, (2) merging closest TC pairs until the number of remaining TCs is 
small enough. We cluster the continuous stream and store the clustering 
snapshots at different timestamps. Since storing the snapshot of every 
timestamp is unrealistic, we use the pyramid time frame (PTF) structure \cite{2} to achieve both excellent space efficiency and high coverage of the stream 
history.

\subsubsection{The Classifier}

We use logistic regression to train a binary classier and judge whether each 
candidate is indeed a local event. We choose logistic regression because of 
its robustness when there is only a limited amount of training data. While 
we have also tried using other classifiers like Random Forest and SVM, we nd 
that the logistic regression classier produces the best result in our 
experiments. The labelled instances for the classier are collected through a 
large-scale experiment on a popular crowdsourcing platform. We will shortly 
detail the annotation process in Section 6.

We analyse the complexity of the candidate classification step as follows. 
As the prediction time of logistic regression is linear in the number of 
features and has $ O(1) $ complexity, the time cost is dominated by the feature 
extraction process. Let NC be the maximum number of tweets in each 
candidate, and $ M $ be the keyword vocabulary size, $ D $ be the latent embedding 
dimension, and $ N_Q $ is the number of tweets in the query window. We need to 
extract the features for all the candidates in the query window. The time 
costs for extracting different features for each candidate event are analyzed as follows: (1) For the temporal unusualness measure, its 
time complexity is $ O(M + NA + D) $ where NA is the maximum number of TCs 
in one snapshot of the activity timeline; (2) For the spatial unusualness 
measure, its time complexity is $ O(M + NQ +D) $; (3) For the temporal ACM 
Transactions burstiness measure, its time complexity is $ O(MNA) $; (4) For the 
spatial burstiness measure, its time complexity is $ O(MNC) $; (5) For the 
static features, the total time complexity is $ O(NC) $.

\subsection{The Online Updater}

In this section, we present the online updater of GeoBurst+. Consider a 
query window $ Q $, let $ Q_0 $ Be the new query window after $ Q $ shifts. Instead of finding the local events 
in $ Q_0 $ from scratch, the online update leverages the results in $ Q $ and updates the event list with 
little cost. If one runs the batch detection algorithm in the updated window 
$ Q_0 $, the candidate generation step will dominate the total time cost in the 
two-step detection process, while the candidate classification step is very 
efficient. Hence, our focus on supporting efficient online detection is to 
develop algorithms that can fast update the geo-topical clustering results 
when the query window shifts from $ Q $ to $ Q_0 $. To guarantee to generate the correct clustering results 
in $ Q_0 $, the key is to nd the new pivots in The new window $ Q_0 $ based on the previous results in $ Q $. Let $ D_Q $ be the tweets 
falling in $ Q $ and $ D_{0Q} $ be the tweets in $ Q_0 $. We denote by RQ the tweets removed 
from $ D_Q $, i.e., $ R_Q = D_Q ..D_{0Q} $; and by $ I_Q $ the tweets inserted into $ D_Q $, 
i.e., $ I_Q = D_{0Q}.. D_Q $. In the sequel, we design a strategy that nds pivots 
in $ D_{0Q} $ by just processing RQ and IQ Recall that, the pivot seeking process 
rst computes the local pivot for each tweet and then performs authority 
ascent via a path of local pivots. So long as the local pivot information is 
correctly maintained for each tweet, the authority ascent can be fast 
completed. The major idea for avoiding ending pivots from scratch is that, 
as $ D_Q $ is changed to $ D_{0Q} $, only some tweets have their local pivots changed. We call them mutated 
tweets, defined as follows.

Definition (Mutated Tweet). A tweet $ d_2 $ $ D_{0Q} $ is a mutated tweet if $ d_1 $ local 
pivot in $ D_{0Q} $ is different from its local pivot in $ D_Q $.

Now the questions are, how do we fast identify the mutated tweets by 
analysing the influence of RQ and $ I_Q $? Our observation is that, for any 
tweet, it can become a mutated tweet only if at least one of its neighbours 
has authority change. Therefore, we take a reverse search strategy to nd 
mutated tweets: (1) First, we identify in $ D_{0Q} $ all the tweets whose authorities have changed. (2) Second, for each 
authority-changed tweet t, we retrieve the tweets that regard t as its 
neighbor and update their local pivots.

\section{EXPERIMENTS}

An Event Radar was implemented to test and simulate our approach as an 
experiment. The setting is a Mac OS laptop with a 1.6GHZ processor and 8GB 
RAM. Event Radar was implemented in MEAN Stack with MongoDB as database and 
Express.js as a server. 

\subsection{Events Visualisation}

Event Radar can visualise all inputs from MongoDB on Google Map and enable 
users to view the events' tag, original tweets, timestamp and the rank score 
of the event. 
\begin{figure}[!htb]
	\includegraphics[width=0.75\linewidth]{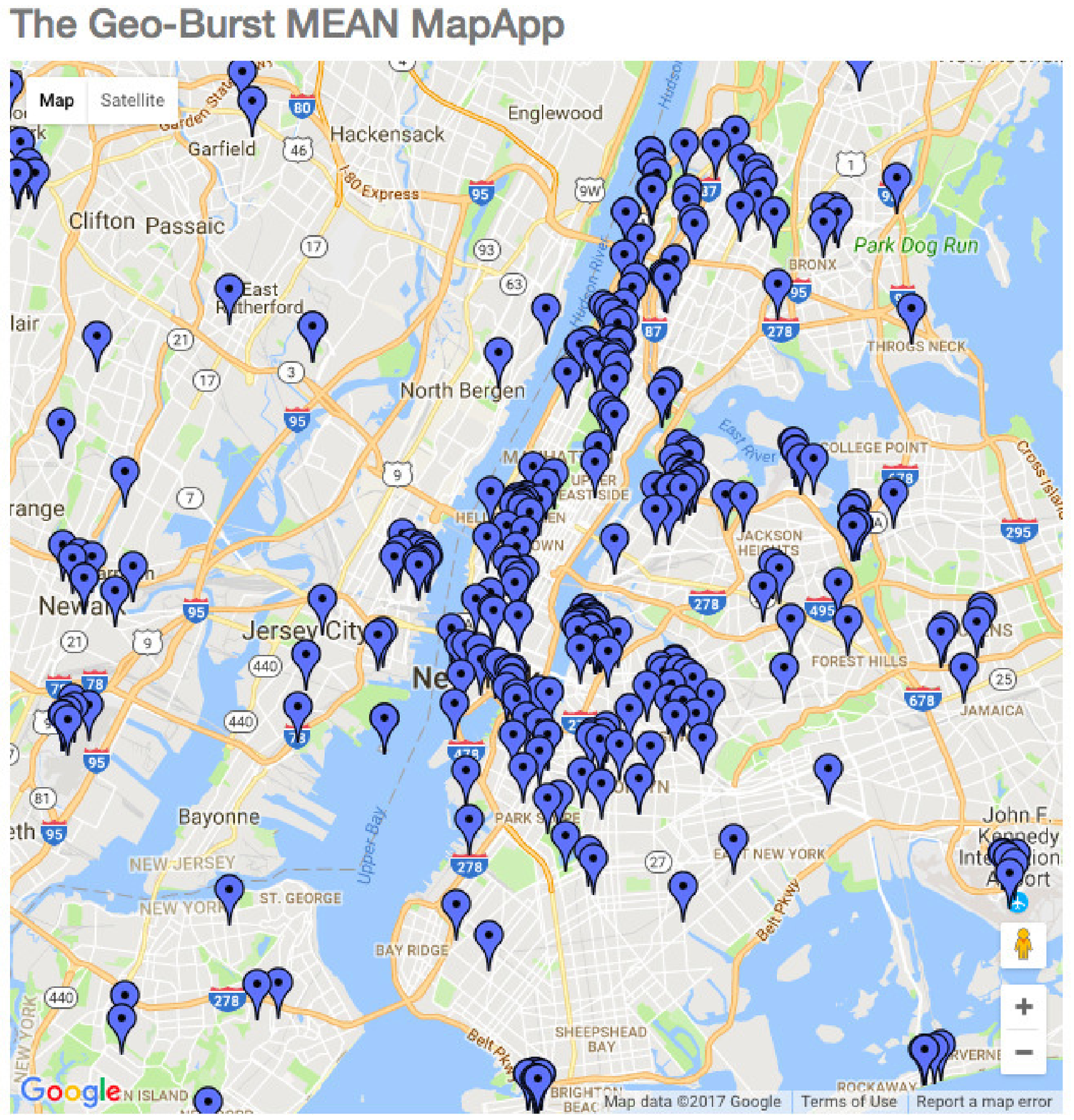}
\end{figure}
\begin{figure}[!htb]
	\includegraphics[width=0.75\linewidth]{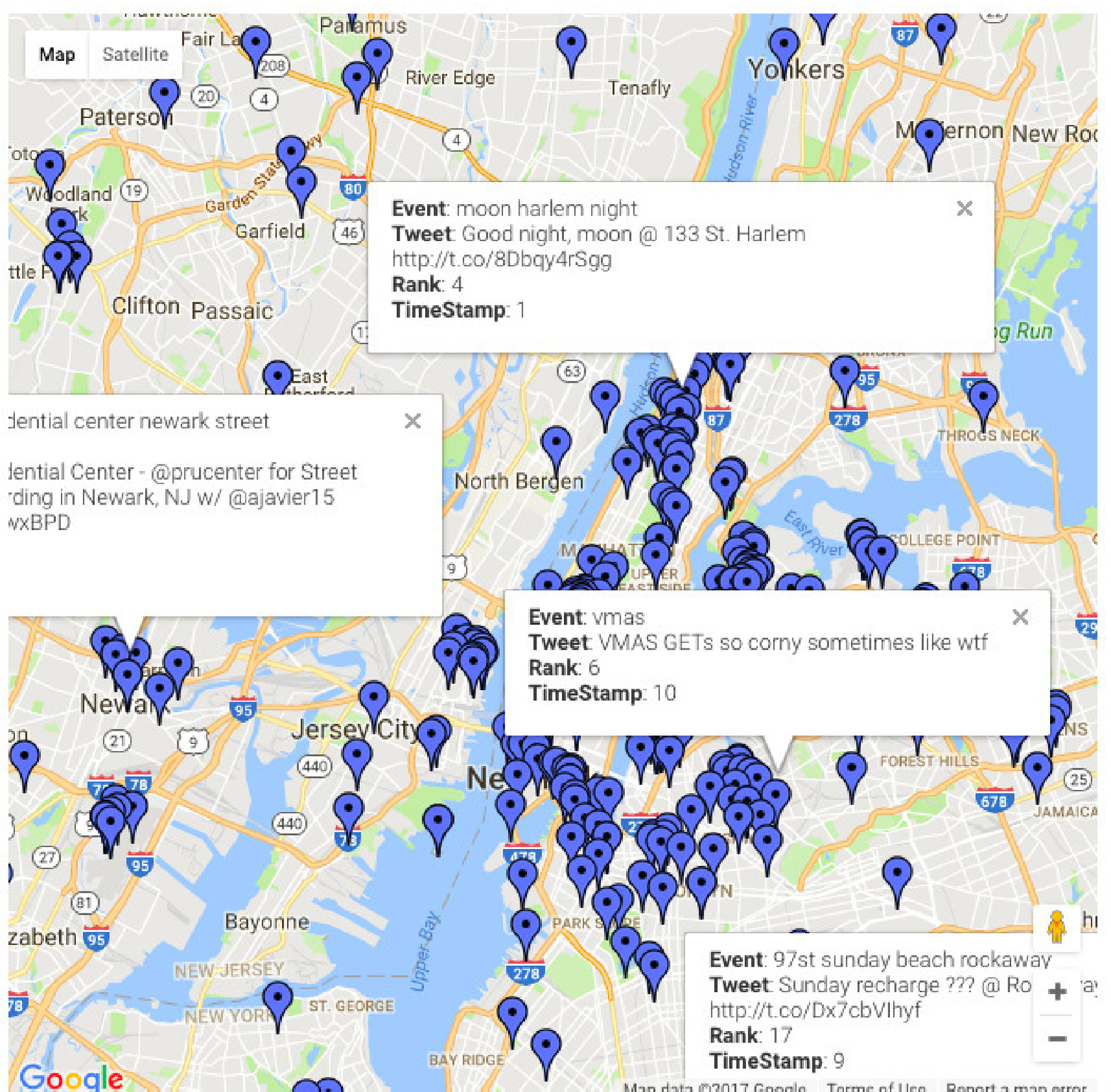}
\end{figure}

\begin{figure*}
	\includegraphics[width=0.75\linewidth]{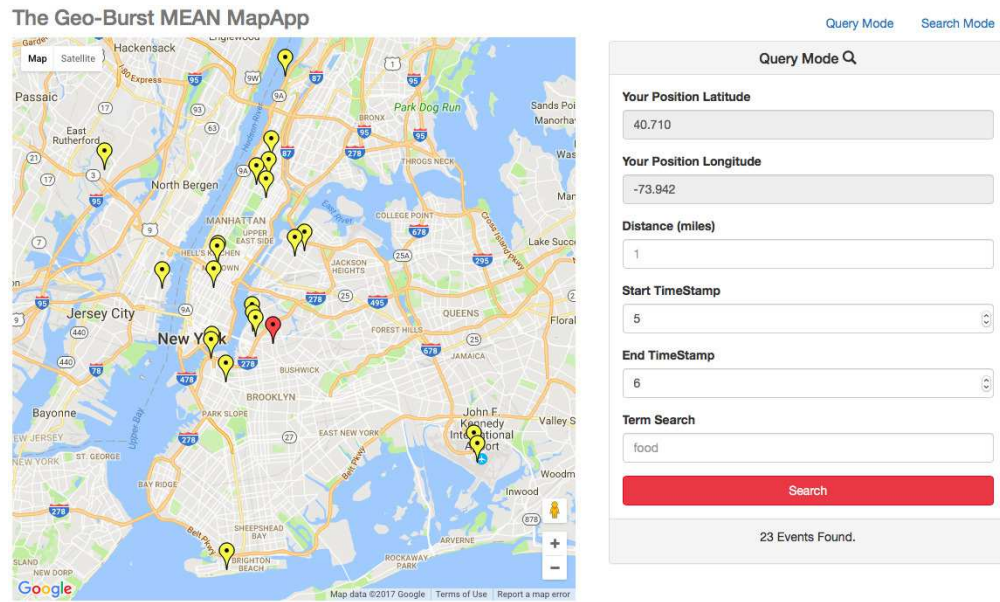}
	\caption{Query of finding the events with a timestamp from 5 to 6. The system returns 23 events.}
\end{figure*}
\begin{figure*}
	\includegraphics[width=0.75\linewidth]{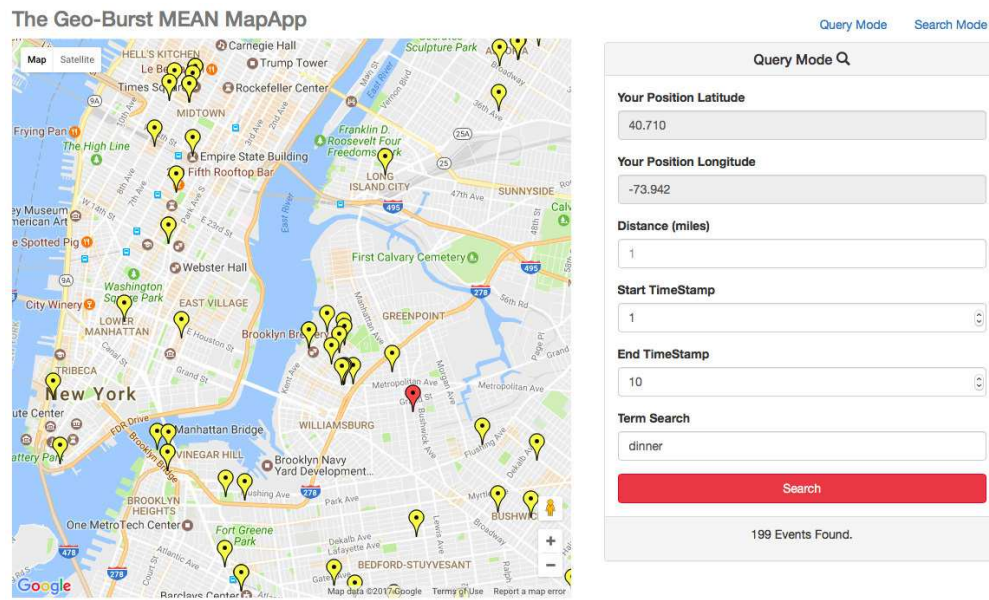}
	\caption{Query to obtain events with a timestamp from 1 to 10, and keyword dinner. 
		The system returns 199 events.}
\end{figure*}

\subsection{Query Mode}

The system also provides a query mode for users to send queries to the 
server to select the desired events, by providing the conditions of event 
tweets' terms, the geospatial distance between the users and events' 
locations, and the timestamp's query windows. 

In summary, Event Radar is a novel approach to providing a web-based 
application for users to view the local events in a given area. 
Additionally, the system contains the query mode for users to search the 
events of their interest. Therefore, such system has a promising perspective 
to be developed as a system for local security authority or press due to the 
reason that it can detect the local events, and update them in the dynamic 
time stream.

\section{CONCLUSION}

We studied the problem of real-time local event detection in geo-tagged 
tweet streams. Event detection aims at finding real-world occurrences that 
unfold over space and time. We mainly implement a website demonstration 
system with an improved algorithm to show the real-time local events online 
for public interests. Our system Event-Radar is not limited to Twitter. 
Rather, any geo-textual social media stream (\textit{e.g.}, Instagram photo tags, 
Facebook posts) can use to extract interesting local events as well. For 
future work, it is interesting to extend Event-Radar for handling the tweets 
that mention geo-entities but do not include exact GPS coordinates.   

We built a demonstration system to visualise the local event detection 
result dynamically. This system consisted of two servers, connected to the 
mongo database. One server is in charge of loading Twitter data from Twitter 
API, constructing of co-occurrence keyword graph, running batch mode to 
generate local event candidates based on geographic impact and semantic 
impact. It ranks the candidate by making vertical comparison across time 
frame and horizontal comparison across all clusters,  finally outputting the 
local event results into Database. Meanwhile, we optimised the original 
project by saving the co-occurrence keyword graph into the database, so that 
when the system restarts, it reloads the graph from the database to save 
sufficient time.  Another server deals with the front end request and 
excellent local event results from the database, sending results to the 
front end to be visualised. \newline

\begin{acks}	
	Special thanks to Chao Zhang, TAs and Prof. Han for giving valuable feedbacks and help during the whole semester.
\end{acks}
\nocite{*}